\documentclass[aps,pra,amsmath,amssymb,twocolumn,showpacs,nofootinbib]{revtex4} 

\usepackage{graphicx}
\usepackage{dcolumn}
\usepackage{bm}
\usepackage{bbold}
\usepackage{pslatex}
\usepackage{epsfig}
\usepackage{color}
\usepackage{nonfloat}

\newcommand{\ket}[1]{\ensuremath{|#1\rangle}}
\newcommand{\bra}[1]{\ensuremath{\langle #1|}}

\newcommand{\be}{\begin{equation}}
\newcommand{\ee}{\end{equation}}
\newcommand{\ba}{\begin{eqnarray}}
\newcommand{\ea}{\end{eqnarray}}

\begin{document}

\preprint{PRE/003}

\title{Hybrid quantum repeater with encoding}

\author{Nadja K. Bernardes$^{1,2}$}%
 \email{nadja.bernardes@mpl.mpg.de}

\author{Peter van Loock$^{1,2,3}$}%
\email{peter.vanloock@mpl.mpg.de}

\affiliation{$^1$Optical Quantum Information Theory Group, Max Planck Institute for the Science of Light, G\"unther-Scharowsky-Str.~1/Bau 24, 91058 Erlangen, Germany}
\affiliation{$^2$Institute of Theoretical Physics I, Universit\"at Erlangen-N\"urnberg, Staudtstr.~7/B2, 91058 Erlangen, Germany}
\affiliation{$^3$ Institute of Physics, University of Mainz, Staudingerweg 7, 55128 Mainz, Germany}



\begin{abstract}
We present an encoded hybrid quantum repeater scheme using qubit-repetition and Calderbank-Shor-Steane codes. For the case of repetition codes, we propose an explicit implementation of the quantum error-correction protocol. Moreover, we analyze the entangled-pair distribution rate for the hybrid quantum repeater with encoding and we clearly identify trade-offs between the efficiency of the codes, the memory decoherence time, and the local gate errors. Finally, we show that in the presence of reasonable imperfections our system can achieve rates of roughly 24 Hz per memory for 20 km repeater spacing, a final distance of 1280 km, and final fidelity of about 0.95.
\end{abstract}

\maketitle

\section{Introduction} 

In 1982, Wootters, Zurek, and Dieks
stated the famous no-cloning theorem \cite{wootters, dieks}. The impossibility to copy an unknown quantum state implies that common procedures used in classical communication to combat channel losses, such as amplification, cannot be used in quantum communication. The problem of distributing entanglement over long distances was then solved in principle with the proposal of quantum repeaters \cite{briegel, dur}. The main idea behind this proposal is to generate entangled pairs in small segments, avoiding the exponential decay the distance, and to use entanglement swapping \cite{zukowski} and entanglement purification \cite{bennett, deutsch} as some of the building blocks of the protocol. Nonetheless, considering the typically probabilistic nature of at least some of these steps (generation, purification, and swapping), the finite decoherence time of the currently available quantum memories turns out to drastically limit the total communication distance. However, according to Ref.~\cite{jiang}, with the help of deterministic quantum error correction (QEC), the initial entangled pairs can be encoded so that all the swapping steps may be executed at the same time. This is different from the original approach of a nested scheme for the repeater (with multiple-round purification and swapping), making the protocol much faster than usual.

There are many different proposals for implementing a quantum repeater, utilizing completely different systems, including heralding mechanisms based on single-photon detection \cite{duan, childress1, childress2, simon, sangouard} and schemes based on bright multiphoton signals. Although in the former schemes, generally, high-fidelity entangled pairs are generated, the latter schemes are usually more efficient, at least for the initial entanglement distribution step. In this work, we will concentrate on the so-called hybrid quantum repeater (HQR) \cite{PvLa, ladd, PvLb}. In this scheme, an entangled pair is initially generated between the electronic spins of two atoms placed in not-too-distant  cavities through an optical coherent state (the so-called ``qubus'').

The main idea of this paper is to apply QEC to a hybrid quantum repeater aiming to improve the scheme against practical limitations such as finite memory decoherence times (relaxing the requirement of perfect memories in our earlier analysis of the unencoded HQR \cite{nadja}) and imperfect two-qubit operations. More specifically, the QEC codes under consideration here are the well-known qubit-repetition and Calderbank-Shor-Steane (CSS) codes \cite{nielsen}. Due to their transversality property, entanglement connection and error correction can be performed with the same set of operations. Our treatment is not restricted to analyzing the in-principle performance of QEC codes for the hybrid quantum repeater, but it also shows how to actually implement an encoded HQR.

In Sec.~\ref{hqr}, we briefly describe the hybrid quantum repeater. The errors affecting the system are presented in Sec.~\ref{errormodel} and the error correcting protocol is described in more detail in Sec.~\ref{qec}. In Sec.~\ref{hqr.rep}, we show how to implement the repetition codes, starting from the general idea and concluding with a proposal for a more practical implementation. A protocol using CSS codes is presented  in Sec.~\ref{hqr.css}. A rate analysis of the hybrid quantum repeater with encoding is presented in Sec.~\ref{rate}. We conclude in Sec.~\ref{conclusion} and give more details of calculations in the appendix.

\section{Hybrid Quantum Repeater} \label{hqr}

A dispersive light-matter interaction provides the essence of the hybrid quantum repeater. This interaction will occur between an electron spin system (i.e., a two-level system or a ``$\Lambda$-system'' as an effective two-level system) inside a cavity and a bright coherent pulse (probe pulse). Although the probe and the cavity are in resonance, both are detuned from the transition between the ground state and the excited state of the atom. More formally, this interaction is described by the Jaynes-Cummings interaction Hamiltonian in the limit of large detuning: $H_{int}=\hbar\chi Z a^{\dagger}a$, where $\chi$ is the light-atom coupling strength, $Z$ is the qubit Pauli-$Z$ operator, and $a$ ($a^{\dagger}$) is the annihilation (creation) operator of the electromagnetic field mode. In practice, this interaction works as a conditional-phase rotation. Considering the two relevant states of the electronic spin as $\ket{0}$ and $\ket{1}$, and a probe pulse in a coherent state $\ket{\alpha}$, we have $U_{int}^q(\theta)\left[(\ket{0}+\ket{1})\ket{\alpha}\right]=\ket{0}\ket{\alpha e^{i \theta/2}}+\ket{1}\ket{\alpha e^{-i \theta/2}}$; $U_{int}^q(\theta)=e^{i(\theta/2) Z a^{\dagger}a}$ is the operator that describes the interaction between the probe and the $q$-th qubit, and $\theta$ represents an effective interaction time, $\theta=-2\chi t$.

First the probe (or qubus) interacts with an atomic qubit $A$ initially prepared in the superposition state $(\ket{0}+\ket{1})/\sqrt{2}$ placed in one of the repeater stations resulting in a qubus-qubit entangled state \cite{PvLa, PvLb, karsten}. Then the qubus is sent to a second qubit $B$ placed in a neighboring repeater station and interacts with this qubit, also initially prepared in a superposition state, this time inducing a controlled rotation by $-\theta/2$. By measuring the qubus (and identifying its state without error, see below), we are able to conditionally prepare an entangled state between qubits $A$ and $B$ which has the following form,
\begin{equation}
F\ket{\phi^+}\bra{\phi^+}+(1-F)\ket{\phi^-}\bra{\phi^-},
\label{finalstate1}
\end{equation}
where $\ket{\phi^\pm}=(\ket{00}\pm\ket{11})/\sqrt{2}$ and $F=[1+e^{-(1-\eta)\alpha^2(1-\cos{\theta})}]/2$, with $\alpha$ real. A beam splitter transmitting on average $\eta$ photons may be used to model the photon losses in the channel. For a standard telecom fiber, where photon loss is assumed to be 0.17 dB per km, the transmission parameter will be $\eta(l,L_{att})=e^{-l/L_{att}}$, where $l$ is the transmission distance of the channel and the attenuation length is assumed to be $L_{att}=25.5$ km. When the optical measurement of the probe pulse corresponds to the quantum mechanically optimal, unambiguous (and hence error-free) state discrimination (USD) of phase-rotated coherent states, an upper bound for the probability of success to generate an entangled pair can be derived, \cite{PvLb}
\be
P_{success}=1-\left(2F-1\right)^{\eta/(1-\eta)}.
\label{P0}
\ee
This bound can be attained, for instance, following the protocol from Ref.~\cite{azuma}. Note that for this type of measurement, there is a trade-off: for large $\alpha$ and hence $F\rightarrow\frac{1}{2}$, we have $P_{success}\rightarrow 1$, as the coherent states become nearly orthogonal even for small $\theta$; whereas for small $\alpha\ll1$ and $F\rightarrow 1$, the coherent states are hard to discriminate, $P_{success}\rightarrow 0$.

Entanglement swapping and purification can also be performed utilizing the same interaction as described above. A two-qubit entangling gate may be employed for both steps. A measurement-free, deterministic controlled-phase gate can be achieved with a sequence of four conditional displacements of a coherent-state probe interacting with the two qubits. The conditional displacements can be each decomposed into conditional rotations and unconditional displacements, so that eventually there is no need for any operations other than those already introduced above \cite{PvLc}. The controlled-phase rotation, single-qubit operations, and measurements are then sufficient tools to implement the standard purification and swapping protocols \cite{deutsch}.

\section{Errors and error models}\label{errormodel}
In the previous section, it was described that the photon losses in the transmission channel cause a random phase-flip error in the initial entangled state. However, considering a more realistic scheme, photon losses will also cause local gate errors, and we should also take into account imperfect memories (i.e., memories with finite decoherence times).

According to Ref.~\cite{louis}, dissipation on \textit{quantum gates} in our scheme will act in a two-qubit unitary operation $U_{ij}$ as
\begin{align}
&U_{ij}\rho U_{ij}^{\dagger}\rightarrow U_{ij}\left[(1-q_g(x))^2\rho+\nonumber\right.\\
&\left.q_g(x)(1-q_g(x))(Z_i\rho Z_i+Z_j\rho Z_j)+q_g^2(x)Z_iZ_j\rho Z_jZ_i\right]U_{ij}^{\dagger},
\label{gateerror}
\end{align}
with 
\be
q_g(x)=\frac{1-e^{-x}}{2},
\ee
the probability that each qubit suffers a $Z$ error, where $x=\frac{\pi}{2}\frac{1-T^2}{\sqrt{T}(1+T)}$; here $T$ is the local transmission parameter that incorporates photon losses in the local gates. Note that this error model is considering a controlled-Z (CZ) gate operation. For a controlled-not (CNOT) gate, Hadamard operations should be included and $Z$ errors can be transformed into $X$ errors.

The errors resulting from the imperfect \textit{memories} are similarly described by a dephasing model, such that the qubit state $\rho_A$ of memory $A$ will be mapped, after a decaying time $t$, to
\be
\Gamma^A_t(\rho_A)=(1-q_m(t/2))\rho_A+q_m(t/2)Z\rho_AZ,
\label{memory.dephasing.1}
\ee
and an initial two-qubit Bell state between qubits $A$ and $B$ will be transformed as \cite{razavi2}
\be
\Gamma^A_t\otimes\Gamma^B_t(\ket{\phi^{\pm}_{AB}}\bra{\phi^{\pm}_{AB}})=(1-q_m(t))\ket{\phi^{\pm}_{AB}}\bra{\phi^{\pm}_{AB}}+q_m(t)\ket{\phi^{\mp}_{AB}}\bra{\phi^{\mp}_{AB}},
\label{memory.dephasing.2}
\ee
where $q_m(t)=(1-e^{-t/\tau_c})/2$ and $\tau_c$ is the memory decoherence time. 

We shall encode our entangled pair in a qubit-repetition code and in a CSS code. The advantage of these codes is that, due to their resemblance to classical codes, the logical operations can simply be understood as the corresponding operations applied upon each physical qubit individually. This permits doing the entanglement connection (swapping) between different repeater stations and the syndrome measurements (for error identification) at the same time, such that the swappings can all be executed simultaneously \cite{jiang}. The correction operations will then be performed only on the initial and the final qubits of the whole protocol. The encoded quantum repeater protocol operates much faster than the non-encoded scheme, and, as a result, still performs well even for rather short memory decoherence times. 

\section{Quantum repeater with error correction}\label{qec}

An $n$-qubit repetition code encodes one logical qubit using $n$ physical qubits in the following way, $\ket{\bar{0}}=\ket{0}^{\otimes n}$ and $\ket{\bar{1}}=\ket{1}^{\otimes n}$. These are the simplest QEC codes, correcting only one type of error (in this case the $X$ error). A more general family of codes that corrects any kind of errors are the CSS codes. A CSS code is constructed from two classical linear codes. Imagine $C_1$ is a linear code that encodes $k_1$ bits in $n$ bits and $C_2$ a linear code 
that encodes $k_2$ bits in $n$ bits, such that $C_2\subset C_1$, and $C_1$ and $C_2^{\bot}$ both correct $(d-1)/2$ errors ($C_2^{\bot}$ is the dual of code $C_2$
). The CSS quantum code is defined as the code encoding $k$ qubits, $k=k_1-k_2$, in $n$ qubits capable of correcting $(d-1)/2$ errors, and is represented by $[n,k,d]$.\footnote{We analyze in this paper only codes with $k=1$. Note that the letter $k$ is used in the rest of the paper for the number of rounds of purification.}
\footnote{Repetition codes in this paper will also be represented by $[n,k,d]$, more precisely, by $[n,1,n]$, since one qubit ($k=1$) is encoded in $n$ physical qubits and the error correcting code will correct $(n-1)/2$ errors.}

According to Ref.~\cite{jiang}, the complete protocol for a quantum repeater with encoding should, in principle, work as follows: first, an encoded Bell pair between two repeater stations is generated. Second, entanglement connection is performed between neighboring stations. Imagine we want to connect the Bell pairs ($A$,$B$) and ($C$,$D$). We should then realize a Bell measurement on the qubits $B$ and $C$. More specifically, this measurement can be performed using a CNOT operation between qubits $B$ and $C$ and a projective $X$-measurement for qubit $B$ and a $Z$-measurement for qubit $C$. For the encoded states, we should be able to perform an encoded version of the Bell measurement. Due to the transversality property of the codes analyzed here, the encoded version of this operation is the same as the operation applied individually for each pair of the $2n$ physical qubits at every repeater station. Provided the system is not noisy and the operations perfect, this would be enough to distribute entanglement over the whole distance. However, of course, this is not a realistic case. The remarkable feature of the encoded scheme \cite{jiang} is now that when performing the entanglement connections, we are able to realize the syndrome measurements at the same time, since we are doing projective measurements on the $2n$ physical qubits. After identifying the error, error correction should be applied and it is guaranteed that the new state is a highly entangled state. All the entanglement connection operations will be performed simultaneously. However, it is important to know exactly which final entangled state is generated. For this purpose, the measurements at the entanglement connection steps will determine the Pauli frame of the final entangled state \cite{jiang}. 

\begin{figure}[h!]
\begin{center}
\epsfig{file=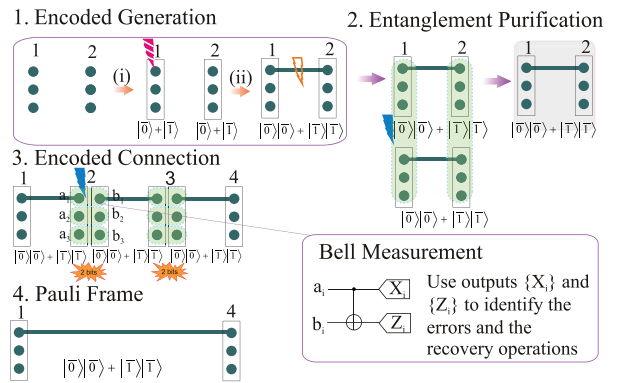,scale=0.6}
\end{center}
\caption{(Color online) Schematic repeater protocol with encoding. In step 1 an encoded entangled pair is distributed. First, at each repeater station there are $2n$ physical qubits. Here $n=3$. In step 1(i), these qubits ($n$ of station 1 and $n$ of station 2) are locally prepared in the encoded state $\ket{\bar{0}}+\ket{\bar{1}}$. By sending and measuring an ancilla qubus state, the state $\ket{\bar{0}\bar{0}}+\ket{\bar{1}\bar{1}}$ is distributed among the two neighboring stations in step 1(ii). Two identical copies of the encoded entangled state are generated, and so applying local operations between each of the $n$ physical qubits in stations 1 and 2 gives a purified entangled encoded state. In step 3, the encoded Bell states are connected, applying Bell measurements individually on each of the $2n$ physical qubits. The outcomes of the Bell measurements on qubits ${a_i}$ and ${b_i}$ are used to  identify the errors and the operations necessary to recover the desired Bell state, or to determine the resulting ``Pauli frame'' (step 4). Errors are represented by a lightning symbol. Red color (striped lightning) indicates when memory errors occur for the first time, orange color (empty lightning) symbolizes imperfect entangled states due to losses in the transmission channel, and blue color (filled lightning) corresponds to errors in the two-qubit gates. }
\label{scheme} 
\end{figure}

The whole protocol, especially a version of it adapted to the use of repetition codes, can now be divided in the following steps: 1) generation  and distribution of the encoded entangled states, 2) purification of the encoded entangled states, 3) encoded entanglement connection, and 4) Pauli frame determination, as illustrated in Fig.~\ref{scheme}. Note that in this version, we first encode and then purify, unlike Ref.~\cite{jiang}. However, as in Ref.~\cite{jiang}, first the codewords are locally prepared and then, with the help of ancilla states, the encoded entangled state is generated. 

There are some peculiarities regarding the different classes of codes and schemes. In the scheme of Jiang \textit{et al.}, first codeword states are locally prepared together with $n$ purified physical Bell states between two repeater stations. An encoded entangled pair is eventually obtained through $n$ pairwise teleportation-based CNOT gates between the local encoded states and the corresponding halves of the Bell states. In contrast, our scheme for the repetition code, as described below, does not require any teleportation-based CNOT gates for the generation of an encoded entangled pair. Consequently, neglecting the CNOT gates necessary for the local codeword generation, while the scheme from Ref.~\cite{jiang} needs in total $4n$ CNOT gates to initially generate a purified encoded entangled pair (for one round of purification), we need CNOT gates only for the purification step. As a result, we use just $2n$ CNOT gates in the preparation of the purified encoded entangled pair. In principle, even these purifications could be done without the use of full CNOT gates \cite{pan, PvL3, denis}. Even more importantly, our protocol for the qubit-repetition code uses only a single lossy channel per encoding block (for any code size $n$), as opposed to the $n$ attenuated Bell pairs in Ref.~\cite{jiang}. Nevertheless, for a version of the protocol based on the CSS codes, as described in detail in Sec.~\ref{hqr.css}, we shall follow a similar strategy to that of Ref.~\cite{jiang}, by teleporting a logical qubit using already prepared Bell states.

The effective logical error probability for each encoding block, after encoding a qubit in an $[n,1,d]$ code and performing syndrome measurement and correction, is
\be\label{Q}
Q_n=\sum_{j=\frac{d+1}{2}}^{n}\left(\begin{array}{c}
n\\
j
\end{array}\right)q_{eff}^{j}(1-q_{eff})^{n-j},
\ee
where $q_{eff}$ is the effective error probability per physical qubit (more details of this will be given below). So, the leading order of errors occurring with probability $q_{eff}$ is reduced to $q_{eff}^{\frac{d+1}{2}}$ through the use of QEC.

Since there are some subtleties regarding the repetition codes and the CSS codes, the two families of codes are analyzed separately below.

\section{Hybrid quantum repeater with repetition code against memory errors}\label{hqr.rep}

Although the repetition code is one of the simplest error-correcting codes, it is not a full quantum error correction code, as it can correct only one type of error. With this in mind, the qubit repetition code will be used here to protect the states against phase-flip ($Z$) errors originating from memory imperfections. For this purpose, the gate errors are considered sufficiently small such that the dominating error is caused by memory imperfections.\footnote{The reason that the $Z$ errors originating from the gate errors are not included in the error correction is that $Z$ and $X$ errors occur with equal probability in the imperfect CNOT gates. In this case, the scheme with the repetition code performs worse than the non-encoded scheme. For example, for the three-repetition code against $X$ errors, the probability of no error will be $(1-q_z)^3(1 - 3 q_x^2 + 2 q_x^3)$. Without encoding, the probability of no error is $(1-q_z)(1-q_x)$. If $q_x=q_z=q$, it is clear that $(1-q)^3(1 - 3 q^2 + 2 q^3)\leq (1-q)(1-q)$.} We produce an encoded entangled state using a qubit-repetition code $[n,1,d]$. For $n=3$, 
$\ket{0}$ is encoded in $\ket{\bar{+}}=\ket{+++}$ and $\ket{1}$ is encoded in $\ket{\bar{-}}=\ket{---}$, where $\ket{\pm}=\frac{\ket{0}\pm\ket{1}}{\sqrt{2}}$. The encoded entangled pairs are connected by applying an encoded Bell measurement between the two half-nodes of the repeater station. This is done by applying pairwise CNOT gates on qubits $\left\{a_i, b_i\right\}$ and by measuring qubits $2a$ in the logical basis $\left\{\ket{\bar{+}},\ket{\bar{-}}\right\}$ and qubits $2b$ in the logical basis $\left\{\ket{\bar{0}},\ket{\bar{1}}\right\}$, as shown in Fig.~\ref{scheme}. The logical computational basis is defined as $\ket{\bar{0}}=\frac{\ket{\bar{+}}+\ket{\bar{-}}}{\sqrt{2}}=\frac{1}{2}(\ket{000}+\ket{011}+\ket{101}+\ket{110})$ and $\ket{\bar{1}}=\frac{\ket{\bar{+}}-\ket{\bar{-}}}{\sqrt{2}}=\frac{1}{2}(\ket{111}+\ket{100}+\ket{010}+\ket{001})$, and it is straightforward to see that by measuring each physical qubit in the $\left\{\ket{0},\ket{1}\right\}$ basis, if the output is an odd number of $\ket{0}$, the logical qubit is in the state $\ket{\bar{0}}$, otherwise, the logical qubit is in the state $\ket{\bar{1}}$. Following this procedure, we will not only connect the encoded entangled states, but we can also identify if an error occurred.

Ignoring the two-qubit gate for the moment, the probability that a logical qubit suffers an error after encoding and applying error correction is given by Eq.~(\ref{Q}), where $d=3$ and $q_{eff}=q_m(t/2)$:
\be
Q_3=q^3_m(t/2)+3q^2_m(t/2)(1-q_m(t/2)).
\ee
For a two-qubit encoded entangled state, the probability that no error occurs is then given by\footnote{Note that this is exactly the same relation that was shown previously in Eqs.~(\ref{memory.dephasing.1}, \ref{memory.dephasing.2}); there $(1-q_m(t))=(1-q_m(t/2))^2+q^2_m(t/2)$.}
\be
\mathcal{P}_3=(1-Q_3)^2+Q_3^2.
\label{effprobsuc}
\ee
The final state after encoding, syndrome measurement, and correction becomes\footnote{Note that for the three-qubit phase-flip code, a logical $\bar{Z}=ZZZ$ operation on the codewords should be seen as a logical $\bar{X}=XXX$ operation on the computational basis.}
\begin{eqnarray}\label{finalstate2}
\mathcal{P}_3\left[F\ket{\bar{\phi}^+}\bra{\bar{\phi}^+}+(1-F)\ket{\bar{\psi}^+}\bra{\bar{\psi}^+}\right]+&&\\
(1-\mathcal{P}_3)\left[F\ket{\bar{\phi}^-}\bra{\bar{\phi}^-}+(1-F)\ket{\bar{\psi}^-}\bra{\bar{\psi}^-}\right],&&\nonumber
\end{eqnarray}
where the encoded versions of the Bell states are represented by $\ket{\bar{\phi}^\pm}=(\ket{\bar{0}}\ket{\bar{0}}\pm\ket{\bar{1}}\ket{\bar{1}})/\sqrt{2}$ and $\ket{\bar{\psi}^\pm}=(\ket{\bar{0}}\ket{\bar{1}}\pm\ket{\bar{1}}\ket{\bar{0}})/\sqrt{2}$. 

Although the encoding protects the original state against memory imperfections, the same does not necessarily happen for the two-qubit gate imperfections. In fact, the effect of these errors may become even stronger in the encoded scheme, affecting, in particular, the purification and swapping steps. We should be very careful here, since the resulting state after the two-qubit interaction will no longer necessarily remain a mixture of Bell states. By having this in mind and the error model in Eq.~(\ref{gateerror}), we are able to estimate the probability of success and the fidelity of the purification and swapping steps. Before getting into details we should remember that, assuming perfect two-qubit gates and an initial state of the form $A\ket{\phi^+}\bra{\phi^+}+B\ket{\phi^-}\bra{\phi^-}+C\ket{\psi^+}\bra{\psi^+}+D\ket{\psi^-}\bra{\psi^-}$ with $\ket{\phi^\pm}=(\ket{00}\pm\ket{11})/\sqrt{2}$ and $\ket{\psi^\pm}=(\ket{01}\pm\ket{10})/\sqrt{2}$, after purification or swapping, the state still has the same form, but with new coefficients given as follows \cite{bennett, deutsch},
\begin{align}\label{Fpure}
&A'_{pur}=\frac{A^2+D^2}{P_{pur}},\quad\quad B'_{pur}=\frac{2AD}{P_{pur}},\nonumber\\
&C'_{pur}=\frac{B^2+C^2}{P_{pur}},\quad\quad D'_{pur}=\frac{2BC}{P_{pur}},
\end{align}
\be
P_{pur}=(A+D)^2+(B+C)^2,
\label{Ppure}
\ee
\begin{align}\label{Fswape}
&A'_{swap}=A^2+B^2+C^2+D^2,\quad B'_{swap}=2(AB+CD),\nonumber\\
&C'_{swap}=2(AC+BD),\quad\quad\quad D'_{swap}=2(BC+AD),
\end{align}
and we will use \cite{nadja}
\be
P_{swap}\equiv 1.
\ee

Considering that the final state will be a complicated mixed state, especially for higher orders of encoding, we include the gate errors by treating these functions in a worst-case scenario. According to this, lower bounds for fidelity and probability of success of purification and for the fidelity of swapping are then given by
\be
P_{pur,lower}(A,B,C,D)=P_{pur}(A,B,C,D)(1-q_g(x))^{4n},
\label{Ppur}
\ee
\be
F_{pur,lower}(A,B,C,D)=A'_{pur}(A,B,C,D)(1-q_g(x))^{4n},
\label{Fpur}
\ee
\be
F_{swap,lower}(A,B,C,D)= A'_{swap}(A,B,C,D)(1-q_g(x))^{2n}.
\label{Fswap}
\ee
For further details, see Appendix A.

We purify previously encoded states, but, since during the entanglement distribution the qubits are already subject to errors, we do error correction before purification. For this, we first apply a Hadamard operation on the qubits changing $\ket{+}$ back to $\ket{0}$, and similarly, $\ket{-}$ back to $\ket{1}$, and then we measure the qubits with the aid of an ancilla state, employing a majority voting. In order to do this, we use a qubus interacting with the atoms in the cavities. The qubus  is measured in the $x$ quadrature and so we can find out if an error occurred which can be corrected. A similar procedure is performed in the implementation of the encoded scheme, which will become clear soon. Since error correction occurs deterministically and locally at each repeater station, this does not affect the generation rates. The purification protocol between the encoded states is very similar to the original version from Ref.~\cite{deutsch}. First, local operations are applied on each physical qubit. At side $A$, the $2n$ physical qubits are subject to the transformation $\ket{0}\rightarrow (\ket{0}+i\ket{1})/\sqrt{2}$ and $\ket{1}\rightarrow (i\ket{0}+\ket{1})/\sqrt{2}$. At side $B$, the $2n$ physical qubits are transformed as $\ket{0}\rightarrow (\ket{0}-i\ket{1})/\sqrt{2}$ and $\ket{1}\rightarrow (-i\ket{0}+\ket{1})/\sqrt{2}$. On both sides CNOT operations are applied transversally on each $n$ physical qubits from the logical control and target qubits. The physical target qubits are measured in the computational basis, and the logical qubits are identified (remember that for the repetition code, an odd number of 0 corresponds to the logical state $\ket{\bar{0}}$ and an even number of 0 refers to the logical state $\ket{\bar{1}}$). Always when the logical qubits measured on both sides coincide, we keep the resulting state and this is a purified encoded entangled state.

The final fidelity of the encoded entangled state, after $k$ rounds of purification and $N-1$ connections (swappings), is given as a lower bound by
\begin{widetext}
\begin{equation}\label{Ftotal}
F_{final}=\underbrace{A'_{swap}(...A'_{swap}(}_{(\log_2{N})-\text{times}}\underbrace{A'_{pur}(...A'_{pur}}_{k-\text{times}}(A_{eff}(F,t_k),B_{eff}(F,t_k),C_{eff}(F,t_k),D_{eff}(F,t_k)))))(1-q_g(x))^{2n((N-1)+2(2^k-1))},
\end{equation}
\end{widetext}
where $A_{eff}(F,t)=\mathcal{P}_n(t)F$, $B_{eff}(F,t)=(1-\mathcal{P}_n(t))F$, $C_{eff}(F,t)=\mathcal{P}_n(t)(1-F)$, $D_{eff}(F,t)=(1-\mathcal{P}_n(t))(1-F)$, $N=L/L_0$ with $L$ the total distance and $L_0$ the fundamental distance between repeater stations, $T_0=2L_0/c$ is the minimum time it takes to successfully generate entanglement over $L_0$, and $c$ is the speed of light in an optical fiber ($2\times 10^8$ m/s). More details can be found in Appendix A. We should be careful when defining the dephasing times $t_k$. We make use of as many spatial resources as need to minimize the required temporal resources, such that the time considered in Eq.~(\ref{Ftotal}), $t_k=(k/2+1)T_0$, is the minimum time it takes for the entanglement distribution and $k$ rounds of entanglement purification to succeed. Notice here that $A_{eff}(F,t_k)$ is smaller than the fidelity that we obtain after entanglement distribution and error correction but before purification, because $t_k\geq T_0$. The probability of success for one round of purification will be estimated as $P_1=P_{pur,lower}(A_{eff}(F,t_1),B_{eff}(F,t_1),C_{eff}(F,t_1),D_{eff}(F,t_1))$. In the case of two rounds of purification, the probability of success will be given by $P_2=P_{pur}(A_{eff}(F,t_2),B_{eff}(F,t_2),C_{eff}(F,t_2),D_{eff}(F,t_2))\times$
\\$P_{pur}(A'_{pur}(A_{eff}(F,t_2),B_{eff}(F,t_2),C_{eff}(F,t_2),D_{eff}(F,t_2)),...,$\\$D'_{pur}(A_{eff}(F,t_2),B_{eff}(F,t_2),C_{eff}(F,t_2),D_{eff}(F,t_2)))(1-q_g(x))^{12n}$. The time spent for the encoding was neglected here, since it will be much shorter than the time spent in classical communication between repeater stations.

How precisely the encoding protocol can be implemented is explained below.
\\

\textbf{Implementation}
\\

\begin{figure}[t b]
\begin{center}
\epsfig{file=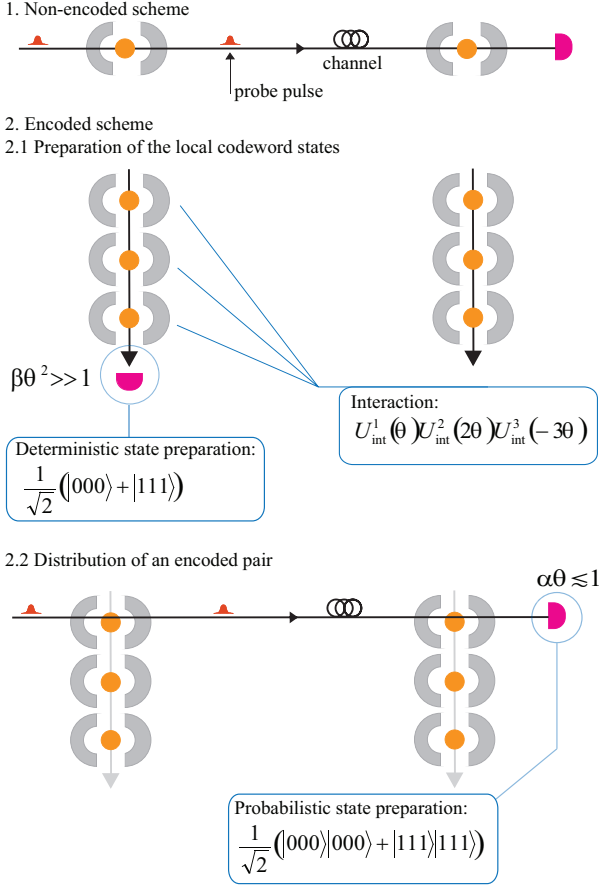,scale=0.45}
\end{center}
\caption{(Color online) Hybrid quantum repeater with repetition code. The qubits are initiated in the state $\left(\ket{0}+\ket{1}\right)^{\otimes 3}$; the normalization factor is omitted. Step 2.1: interacting the qubits with a coherent state $\ket{\beta}$. By measuring $\ket{\beta}$, the state $\ket{000}+\ket{111}$ is prepared. Step 2.2: a probe state $\ket{\alpha}$ interacts with \textit{only one qubit} at each repeater station. By measuring the qubus, the entangled encoded state $\ket{000}\ket{000}+\ket{111}\ket{111}$ is prepared.}
\label{fig} 
\end{figure}

The implementation for the repeater with encoding, omitting purification, can be described as follows. In the unencoded scheme, a probe beam interacts with two qubits at two neighboring repeater stations. With encoding, it is crucial to observe (see below) that we may still use only one probe beam, however, $n$ qubits per half node are needed. Initially, the qubits are all in the state $((\ket{0}+\ket{1})/\sqrt{2})^{\otimes n}$. It is important for the encoded scheme that the generation of the locally created encoded states occurs deterministically, \footnote{This is similar to Ref.~\cite{jiang} where first GHZ states are produced locally, which are then teleported into the initially created nonlocal Bell pairs. Note that in Ref.~\cite{jiang}, the purification of these initially distributed Bells pairs is also made near-deterministic by using sufficiently many temporal and spatial resources. We shall follow a similar strategy, but in our case for the initial distributions, see Sec.~\ref{rate}.} since otherwise the whole protocol would again become too slow. Through interaction of the qubits with a coherent state with sufficiently large amplitude, $\beta\gg1$ with $\beta$ real, it is possible to prepare the $n$-qubit state $(\ket{0}^{\otimes n}+\ket{1}^{\otimes n})/\sqrt{2}$, for example, employing homodyne measurements. This works because the interaction between qubits and qubus (probe) functions as a controlled-phase rotation and we are, in principle, able to deterministically distinguish between the phase-rotated components of $\ket{\beta}$ by measuring the $x$ quadrature (that is perpendicular to the direction of the phase rotation). For $\beta\gg1$, this can be even achieved in an almost error-free fashion. By preparing the qubits in this way, the transmitted qubus beam (between two stations) will interact only with one qubit pair from the chains of $n$ qubits. More specifically, let us take a look at the 3-qubit repetition code as illustrated in Fig.~\ref{fig}. The qubits are initiated in the state $\left(\frac{\ket{0}+\ket{1}}{\sqrt{2}}\right)^{\otimes 3}$. As shown in step 2.1, this state interacts with a coherent state $\ket{\beta}$, and this interaction is described by
\begin{align}
&U_{int}^1\left(\theta\right)U_{int}^2\left(2\theta\right)U_{int}^3\left(-3\theta\right)\left[\left(\frac{\ket{0}+\ket{1}}{\sqrt{2}}\right)^{\otimes 3}\ket{\beta}\right]=\nonumber\\
&\frac{1}{2\sqrt{2}}\left[\left(\ket{000}+\ket{111}\right)\ket{\beta}+\ket{001}\ket{\beta e^{3i \theta}}+\ket{010}\ket{\beta e^{-2i \theta}}+\right.\nonumber\\
&\left.\ket{100}\ket{\beta e^{-i \theta}}+\ket{110}\ket{\beta e^{-3i \theta}}+\ket{101}\ket{\beta e^{2i \theta}}+\ket{011}\ket{\beta e^{i \theta}}\right].
\end{align}
By measuring the $x$ quadrature of the probe beam, the state $\frac{\ket{000}+\ket{111}}{\sqrt{2}}$ is deterministically generated up to a known phase shift \cite{nemoto} and local bit flip operations. In the next step 2.2, a probe state $\ket{\alpha}$ interacts with only \textit{one qubit} at each repeater station. This time, after performing a USD measurement on the qubus, as was explained in Sec.~\ref{hqr}, the entangled encoded state 
\be
\frac{\ket{000}\ket{000}+\ket{111}\ket{111}}{\sqrt{2}}
\ee
is prepared.\footnote{Note here that there is an important difference between the state preparation in step 2.1 and in step 2.2. In the first step, the coherent state $\ket{\beta}$ interacts with cavities that are locally positioned next to each other, and by using a sufficiently bright beam, $\beta\gg1$, homodyne measurements in the $x$ quadrature will be enough to deterministically prepare the state up to a known phase shift and local bit flip operations. In the second step, the probe beam interacts with two cavities spatially separated from each other. In this case, the effect of photon losses in the channel depends on the amplitude of the beam, and we cannot make $\alpha$ arbitrarily large. Consequently, the generation of the entangled state must become non-deterministic.} To prepare an encoded entangled state in the conjugate basis, we just have to apply Hadamard operations on the physical qubits immediately after the local codeword state has been produced and the probe beam has interacted with one of the qubits. We assumed here that the codeword state at side $B$ is prepared only at the very moment when the probe qubus arrives at this side, thus avoiding memory dephasing during the transmission time. For larger codes ($n>3$), similar sequences of interactions can be found. However, considering the typical size of $\theta$ and the number of interactions, it will be more practical to use more than one local qubus beam for the preparation of the encoded state. For more details, see Appendix B.

Although the scheme presented here has a fairly simple description, its experimental implementation can be technologically challenging. For an almost error-free and deterministic scheme, there is the constraint $\beta\theta^2\gg1$,\footnote{If we did allow for an almost error-free but probabilistic scheme, we would have $\beta\theta\gg1$ using $p$-quadrature measurements.} which for small phase shifts of $\theta\sim10^{-2}$ requires bright beams or even ultrabright beams (i.e., pulses with mean photon number larger than $10^8$). Notice that the probability of error caused by the nonorthogonality of the coherent states with finite amplitude is orders of magnitude smaller than the other errors considered here ($P_{error}<10^{-5}$ for $\beta\theta^2>9$ \cite{nemoto}) and will be neglected in our analysis. Moreover, in our scheme the main contribution of losses is through the transmission channel; losses that happen in the interaction between the qubus state $\ket{\beta}$ and the atomic qubits are strictly local and hence will be neglected here.\footnote{Note that unlike these encoding and entangling steps, for the purification and swapping, local losses will be included, as the latter require (in the present protocol) full CNOT gates which are most sensitive to the local losses.}
Note that prior to entanglement purification, the only remaining elements which are probabilistic and which contribute to the infidelity of the encoded pairs are the postselection of the single qubus beam and its lossy transmission, respectively. The postselection is then achieved through a USD measurement (Fig.~\ref{fig} and Eq.~(\ref{P0})). In the protocol of Ref.~\cite{jiang}, in contrast, first, all of the $N_0>n$ Bell pairs are subject to fiber attenuations through their channel transmissions (see App. D of Ref.~\cite{jiang}), prior to their purifications and the encoding steps (in precisely this order, as opposed to that of our protocol with first encoding, second, transmission, and third, purification). As a result, in our scheme, we minimize the effect of the lossy channel transmissions and the corresponding need for entanglement purification.

\section{Hybrid quantum repeater with CSS code against memory and gate errors}\label{hqr.css}

One important property of a CSS code is that the encoded version of many important gates can be implemented transversally, i.e., the encoded version of the gate is simply the same gate applied individually to each physical qubit in the code. For example, the encoded version of the $X$ operation, i.e., the logical $X$, is represented by $\bar{X}$ with $\bar{X}=X^{\otimes n}$. Not only does the $X$ operation have a transverse implementation, but so do any Pauli and Clifford gates such as $Y$, $Z$, Hadamard, CNOT, and CZ. These are exactly the operations we will need in our scheme. \footnote{An example of a gate that cannot be implemented transversally in a CSS code is the $\pi/8$ gate \cite{nielsen}, which is a non-Clifford gate.} 

Moreover, the encoded version of some measurements (for instance, in the eigenbasis of $X$, $Y$, and $Z$) have also a transverse implementation. Consider a measurement in the $\bar{Z}$ basis on an arbitrary encoded state $a\ket{\bar{0}}+b\ket{\bar{1}}$. The resulting state will be with probability $\left|a\right|^2$ the state $\ket{\bar{0}}$ and with probability $\left|b\right|^2$ the state $\ket{\bar{1}}$. Similarly, if we measure all the $n$ qubits in the $Z$ basis, we obtain the Hamming weight \footnote{In the present context, the Hamming weight is the number of physical qubits that are different from 0 in a state. The CSS codes have the nice property that if one of the codewords, for example $\ket{\bar{0}}$, has an even Hamming weight, then the complementary codeword, in this example $\ket{\bar{1}}$, has odd Hamming weight.} of the state and, consequently, we get the correct result.

Exploring then the possibility of the transverse implementation of encoded operations and measurements, the protocol for the quantum repeater with encoding using CSS codes can be executed similarly to what was explained above. The main difference now is the order of the encoding and the purification steps. For the repetition codes, purification occurs after encoding. The same procedure could be applied for the CSS codes, but in this case the distance $d$ is usually smaller than the number of qubits $n$, and therefore, when more than $\frac{d-1}{2}$ errors occur, the state is not necessarily defined in the codeword space anymore. This causes the purification protocol to work extremely inefficiently. Consequently, for the CSS codes we follow the same strategy as in Ref.~\cite{jiang}: the encoded entangled state is prepared by teleporting a logical qubit generated locally using $n$ already prepared purified Bell states distributed between the repeater stations.

Assuming that the error probabilities $q_g$, $q_m$, and $(1-F)$ are sufficiently small, as in Ref.~\cite{jiang}, we estimate an effective error probability per physical qubit as
\be
q_{eff}=3q_m(t'_k/2)+2q_g(x)+(1-F).
\label{qeff}
\ee
If purification occurs, $(1-F)$ will be replaced by $(1-F_k)$. Here, $F_k$ is the fidelity after $k$-rounds of purification using the initial state of Eq.~(\ref{finalstate1}), the purification protocol from Ref.~\cite{deutsch}, and the gate error from Eq.~(\ref{gateerror}); an explicit formula and further details are presented in Appendix C. We further exploit that the memory decaying time is $t'_k=(k+1)T_0/2$, assuming that for the distribution of the entanglement (and for purification) the qubits suffer memory dephasing just during the time it takes for classical communication of a successful distribution (purification) event.

The effective logical error probability is given by combining Eqs.~(\ref{Q}, \ref{qeff}). The final fidelity is given by
\be
F_{final}=(1-Q_n)^{2N}.
\ee

Although we do not propose an explicit implementation for the CSS codes, the generation of the codeword states from the CSS codes (in the form of cluster states), using weak nonlinearities similar to those employed in the HQR, was proposed in Ref.~\cite{louis2}. However, that scheme, in its most practical manifestation, is probabilistic. This probabilistic feature will drastically decrease the generation rates and require longer-lasting memories (suppressing the benefit of the encoding against memory errors). Instead, in Ref.~\cite{lin1}, the codewords are created in a deterministic fashion using a similar hybrid system. However, the codeword cluster states generated in this proposal are in fact photonic states, which work badly as a memory. Nonetheless, in principle, a similar approach appears feasible also in the present context of CSS encoding for the HQR using atomic memories. We leave a detailed proposal of an explicit implementation of a CSS HQR for future research.

\section{Rate analysis}\label{rate}

Complementary to our analysis in Ref.~\cite{nadja}, the pair creation rates will now be calculated, assuming, as in Refs.~\cite{jiang, perseguers, razavi2, munro2}, that there are sufficiently many initial resources, such that it is (almost) guaranteed that at least one entangled pair will be successfully generated between two neighboring repeater stations. In other words, for instance, for the repetition codes, we assume $s\gg 1$, where $s$ is the number of memory blocks in each half repeater station. In every block there are $n$ memory qubits, conditionally prepared in the state $\frac{\ket{\bar{0}}+\ket{\bar{1}}}{\sqrt{2}}$. To give an example, in Fig.~\ref{fig}, the case of one block, $s=1$, and three physical qubits, $n=3$, is shown. Assuming that we have to distribute entanglement only for the top physical qubits of the blocks, the average number of encoded entangled pairs generated at time $T_0$ will be $sP_0$, where $P_0$ is the probability of success of generating an entangled pair, in our case given by Eq.~(\ref{P0}). The rate of generating an encoded entangled pair would be given then by $\frac{sP_0}{T_0}$. The rate of successfully generating an encoded entangled pair per each of the $sn$ memories employed in every half node of the repeater is then $\frac{P_0}{nT_0}$. Since the swapping step is taken to be deterministic, this can be considered also as the rate of successful generation of an encoded entangled pair over the total distance $L$ (without purification).

For the CSS codes, let us use $s'$ as the total number of physical qubits available at each half node of the repeater station which are involved in the distribution of the entangled states. This number of entangled pairs is on average given by $s'P_0$. Since for each encoding block we need at least $n$ entangled pairs to teleport the logical qubits, the average number of encoded entangled pairs is calculated as $\frac{s'P_0}{n}$. The rate to generate an encoded pair is then given by $\frac{s'P_0}{nT_0}$. Note that we use $s'$ both for the ``flying" and ``stationary" (memory) qubits, such that the rate to generate an encoded entangled pair per memory can be written as $\frac{P_0}{nT_0}$. This is in fact an overestimation of the number of stationary memories, since not all physical qubits involved need to be memory qubits.

To summarize, the rate of successful generation of an entangled pair over a total distance $L$, divided into segments $L_0$, per each memory employed in every half node of the repeater, for both repetition and CSS codes, is given by
\be
R_n=\frac{P_0}{nT_0}.
\ee  
Notice here that for the scheme without purification, the memory and gate errors will affect the final fidelity of the entangled state, but will have no direct impact on the rates.

Depending on the application aimed at for the resulting large-distance entangled pair, purification should be included in the QEC protocol. The rates including purification are described by
\be
R_{pur,n}=\frac{P_0P_k}{n2^k(k/2+1)T_0},
\label{ratepur}
\ee
where $P_k$ is the probability of success for the $k$-th purification step. For the repetition code, $P_k=P_{pur,lower,k}$, defined in Eq.~(\ref{Ppur.rep}) in Appendix A. In the case of CSS codes, $P_1$ is defined using Eq.~(\ref{ppur}); for more rounds of purification it is possible to deduce more general expressions for $P_k$ with the help of the results from Appendix C. The factor $2^k$ appears, because for each round of the purification, already initially twice as many entangled pairs are necessary. The time it takes to produce an encoded purified entangled pair is $(k/2+1)T_0$; $T_0$ is the time it takes to distribute successfully the entangled pairs and $k/2$ is the time spent to communicate that purification succeeded. Compared to the time spent on classical communication between repeater stations, the time needed for the local operations is much shorter, and so these operation times are neglected here. Since all the swappings happen at the same time, purification will occur only at the first nesting level, as in Ref.~\cite{nadja}. Let us now discuss the rates that we obtained.

\begin{figure}[h!]
\epsfig{file=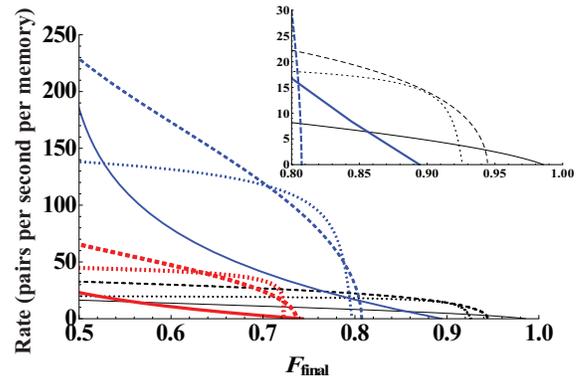, scale=0.5}
\caption{(Color online) Rates for a HQR without purification (solid line), with one round of purification (dashed line), and two rounds of purification (dotted line) in the first nesting level with $L=1280$ km, $L_0=20$ km, $\tau_c=0.1$ s, and $1-T=0.1\%$. Blue curves are the scheme without encoding, red (thick) curves for the scheme with encoding in the $[3,1,3]$ code, and black (thin) curves for the scheme with encoding in the $[7,1,3]$ code.}
\label{fig2a}
\end{figure}

First, in Fig.~\ref{fig2a}, the rates are shown to generate an entangled pair for a total distance of $L=1280$ km, $L_0=20$ km, without purification, and with one and two rounds of purification. Here we considered an imperfect memory with decoherence time $\tau_c=0.1$ s,\footnote{Currently experimentally available memory times are of the order of ms for electronic spins and s for nuclear spins.} while the parameter of local losses in the CNOT gates is $1-T=0.1\%$. We compared the performance of various schemes, namely, encoding with the three-qubit repetition code $[3,1,3]$, and with the Steane code $[7,1,3]$, and without encoding. We will stick in the rest of our analysis to two rounds of purification, as typically this turned out to be the best approach; however, as can be seen in Fig.~\ref{fig2a}, it is not always the best choice. We find that in order to achieve $F_{final}>0.9$, encoding is absolutely necessary in this parameter regime, and that a suitable code is the Steane code.

In Fig.~\ref{fig3}, we plotted the rates for $L=1280$ km, $L_0=20$ km, and the following codes: 3-qubit repetition $[3,1,3]$, 7-qubit repetition $[7,1,7]$, 51-qubit repetition $[51,1,51]$, Steane $[7,1,3]$, Bacon-Shor $[25,1,5]$, and Golay $[23,1,7]$; and for comparison, also the non-encoded scheme. We plotted the rates for different values of $\tau_c$ and $T$: $1-T=0.1\%$ (top), $1-T=0.01\%$ (bottom), $\tau_c=0.01$ s (left), $\tau_c=0.1$ s (center), and $\tau_c=1$ s (right). As expected, repetition codes (which cannot correct gate errors) perform better when the gate errors are sufficiently small ($1-T=0.01\%$). We also observe that for $\tau_c\leq 0.01$ s, the CSS codes have a very bad performance. However, $\tau_c=0.1$ s, even with $1-T=0.1\%$, is already enough to achieve high final fidelities ($F_{final}>0.9$) using the CSS codes. It is interesting to notice that for the parameters presented here, the $[3,1,3]$ code always performs better than the other repetition codes. This can be understood by noting that the bigger the repetition code is, the more susceptible it is to gate errors. We would like to mention that even if we allow for decoherence times of the order of $\tau_c=10$ s, the HQR can still not afford gates with loss parameter of the order of $1-T=1\%$. 
 In addition, a scheme with $1-T=0.01\%$ performs almost identically to a scheme with perfect gates, $1-T=0$. 

It is clear that there are trade-offs between the efficiency of the codes versus the values of decoherence time and the local gate error parameter. As usually observed in QEC schemes, to make the code more complicated and bigger (i.e., use larger spaces and bigger circuits) would in principle suppress the errors more effectively; however, all the extra resources and gates are also subject to errors; so one not only reduces the existing errors, but also introduces new sources of errors. Considering $F_{final}=0.95$, for $\tau_c=0.01$ s and $1-T=0.01\%$, the three-repetition code (see Fig.~\ref{fig3}, left bottom) achieves a rate of about 24 pairs per second per employed memory qubit. For $\tau_c=0.1$ s and $1-T=0.1\%$, and the same final fidelity, the $[23,1,7]$ code (see Fig.~\ref{fig3}, center top) can achieve a rate of about 6 pairs per second per memory. However, for $\tau_c=1$ s and $1-T=0.1\%$, the $[7,1,3]$ code (see Fig.~\ref{fig3}, right top) can achieve rates of about 14 pairs per second per memory. Note that the final fidelities presented here are those exactly obtained at the time when the entangled pair was distributed over the entire distance $L$. Consequently, the dephasing errors due to memory imperfections will continue to degrade the fidelity whenever the final pair is not immediately consumed and used in an application.

\begin{widetext}
\begin{center}
\begin{figure}[t]\label{ratememorytotal}
\epsfig{file=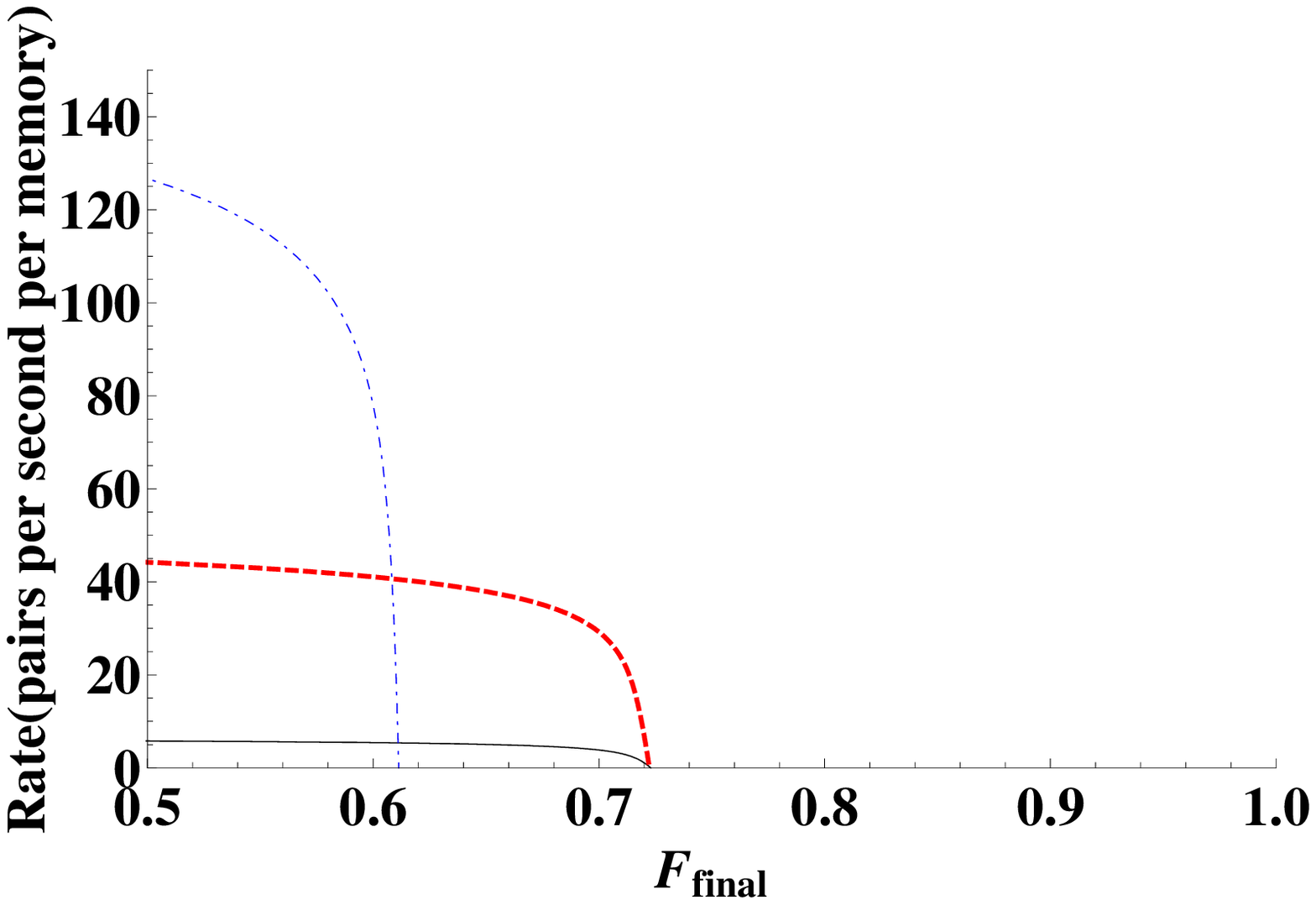,scale=0.36}  \epsfig{file=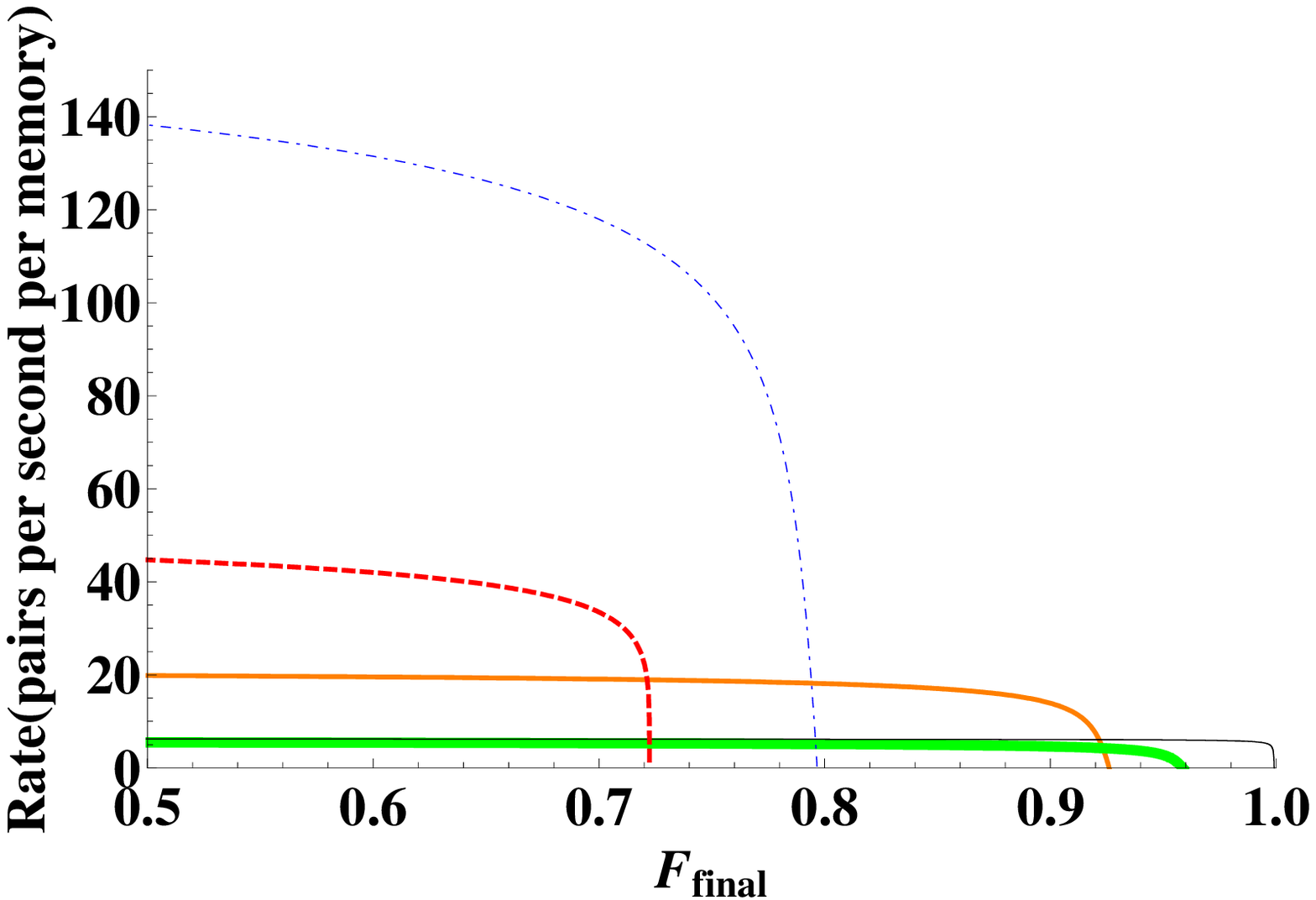,scale=0.36}  \epsfig{file=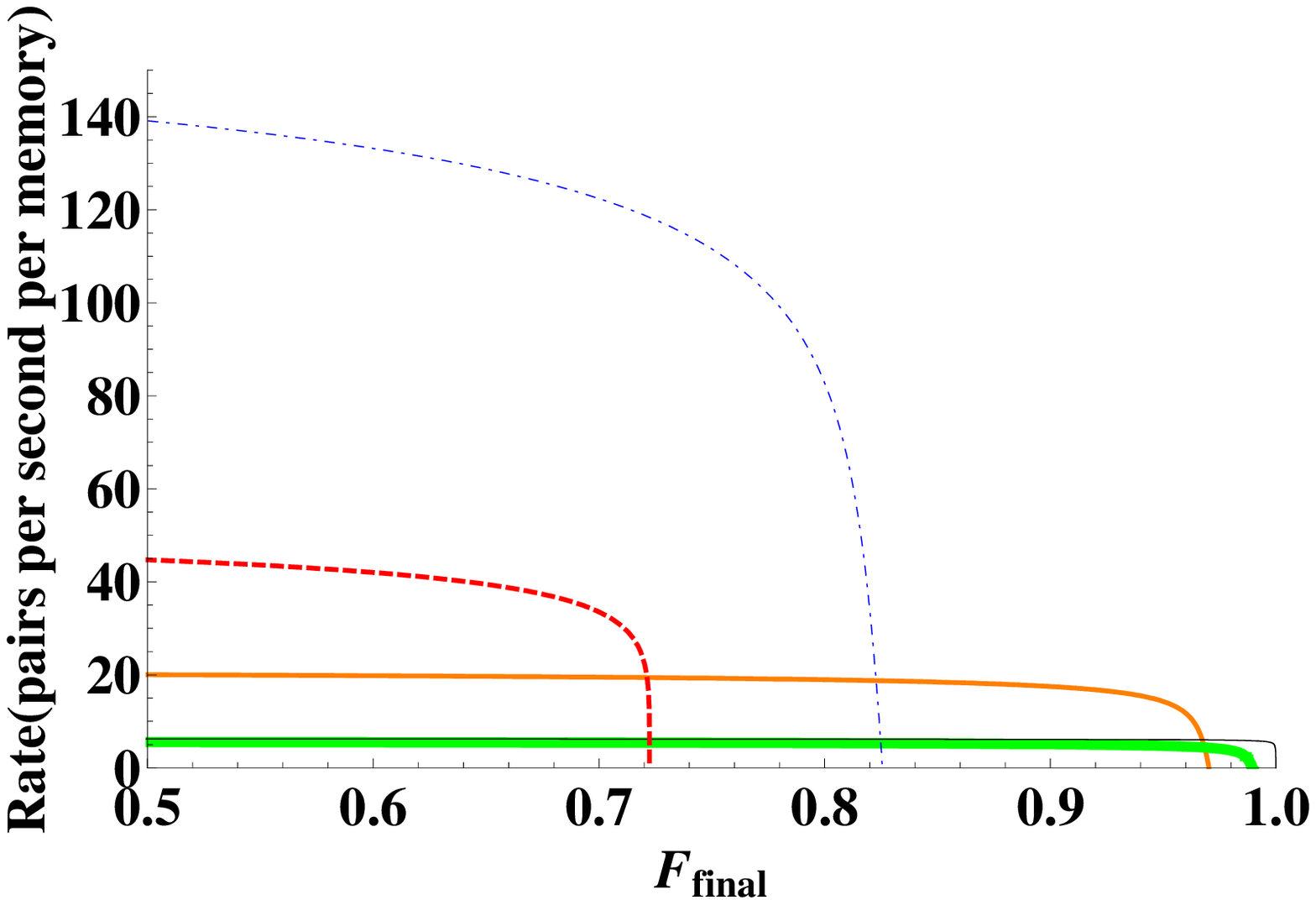,scale=0.36}\\
\epsfig{file=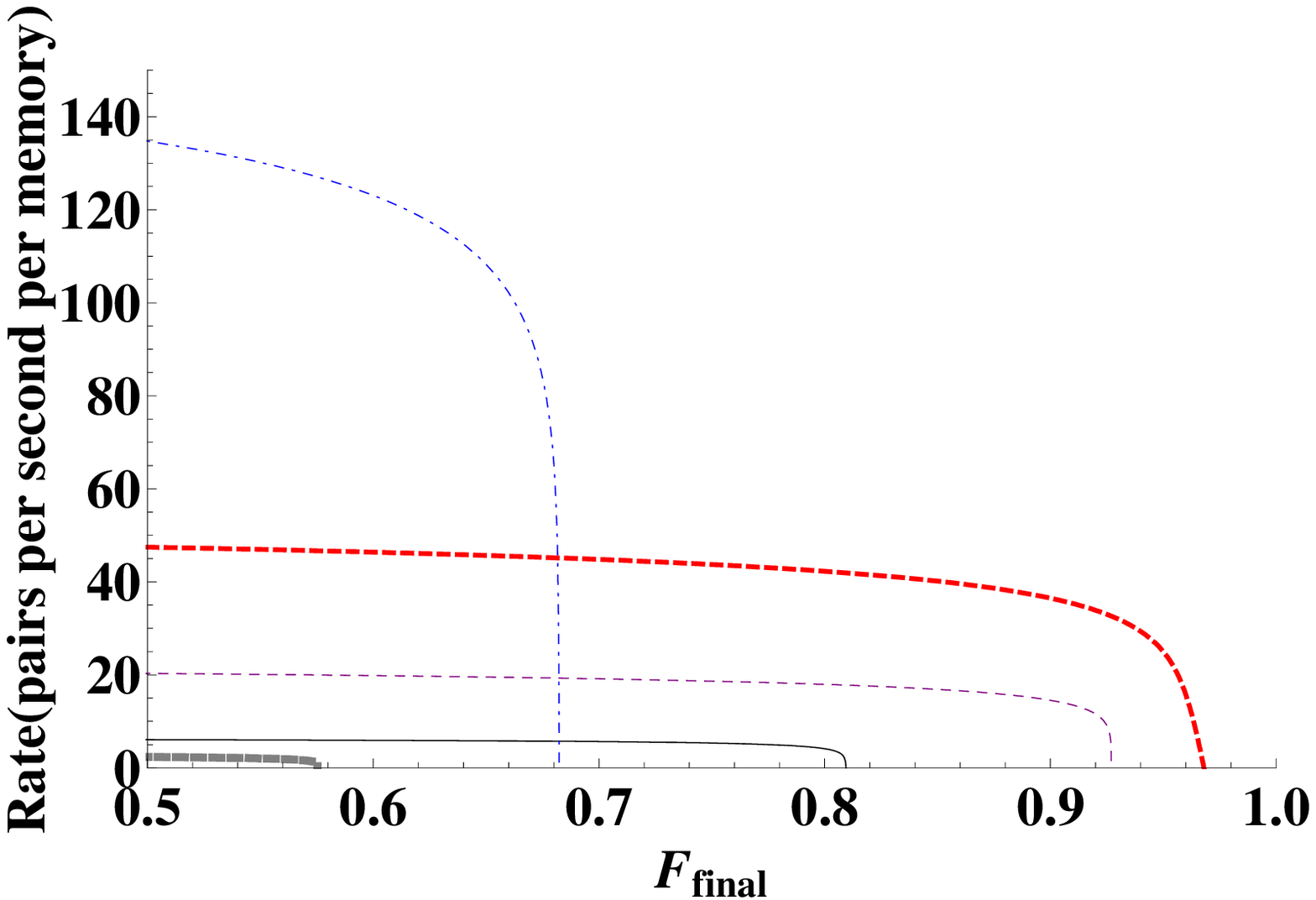,scale=0.36}  \epsfig{file=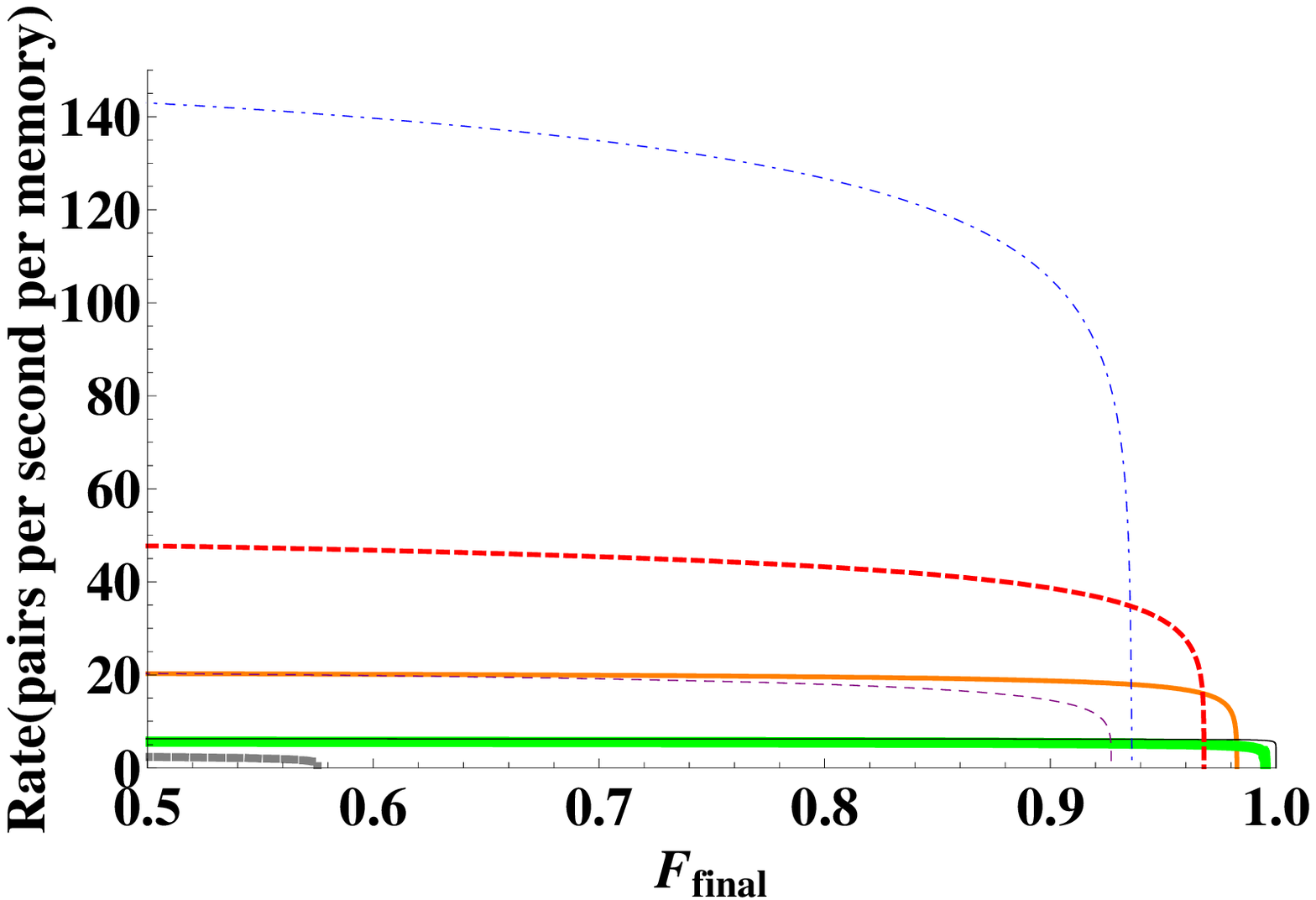,scale=0.36}  \epsfig{file=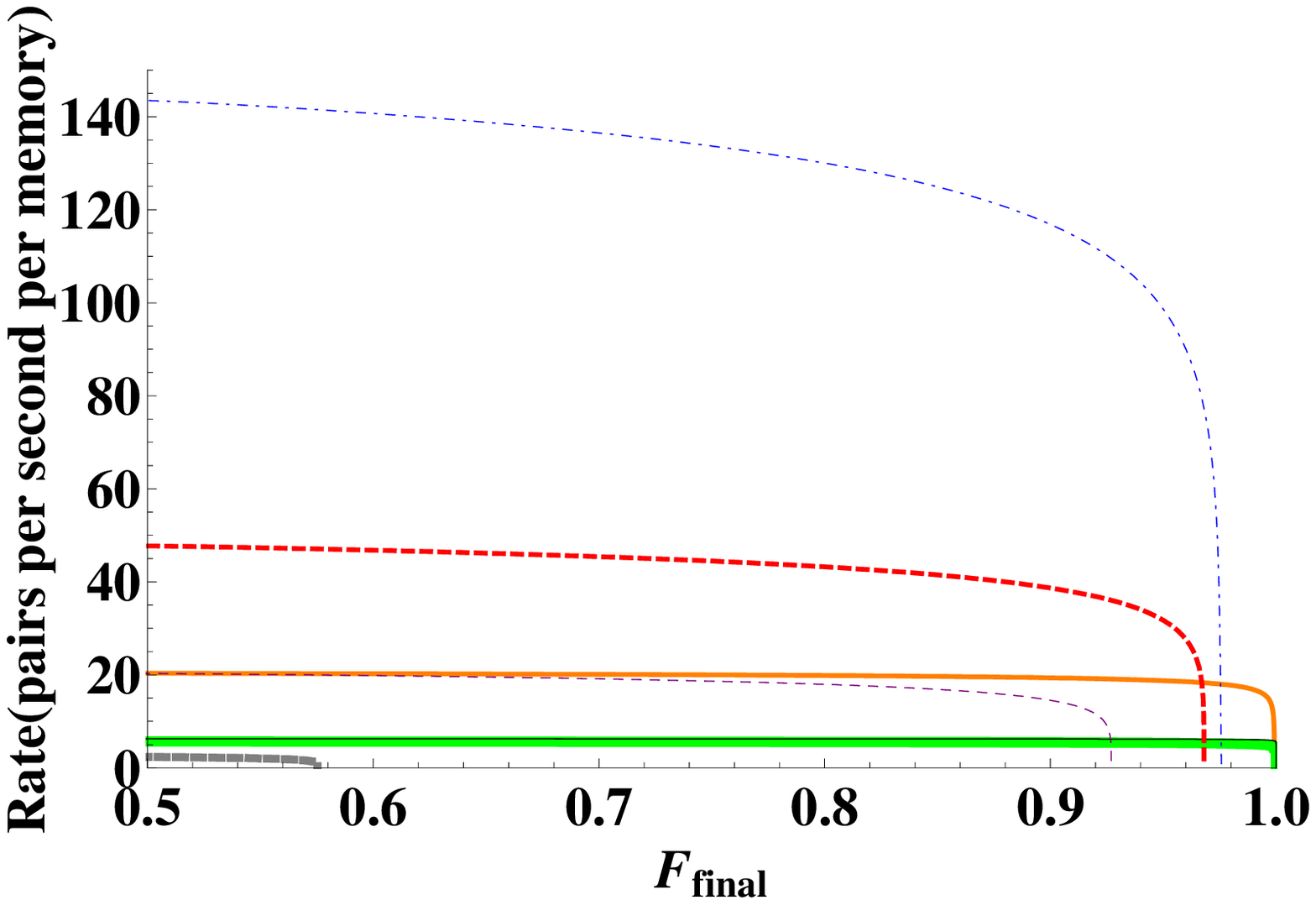,scale=0.36}\\
\caption{(Color online) Rates for a HQR with two rounds of purification in the first nesting level with $L=1280$ km, $L_0=20$ km, $\tau_c=0.01$ s (left), $\tau_c=0.1$ s (center), $\tau_c=1$ s (right), $1-T=0.1\%$ (top), and $1-T=0.01\%$ (bottom). Blue dot-dashed (thin) line is for non-encoded, red dashed line for the [3,1,3] code, purple dashed (thin) line for the [7,1,7] code, gray dashed (thick) line for the [51,1,51] code, orange solid line for the [7,1,3] code, black solid (thin) line for the [23,1,7] code, and green solid (thick) line for the [25,1,5] code.}
\label{fig3}
\end{figure}
\end{center}
\end{widetext}

\section{Conclusion}\label{conclusion}

We presented here an explicit protocol for a hybrid quantum repeater including the use of QEC codes in the presence of imperfect quantum memories and local gate errors. We showed for the case of repetition codes how encoded states can be generated utilizing the same interactions as for the unencoded scheme. Moreover, we calculated the entanglement generation rates and, to properly compare the different schemes, we computed here the rates per memory qubits. We showed that our system, with [23,1,7], with reasonable imperfections, can achieve rates of 6 pairs per second per memory with final fidelities of about $F=0.95$ for a repeater spacing of $L_0=20$ km, a final distance of $L=1280$ km, local gate errors of $1-T=0.1\%$, and a decoherence time of $\tau_c=0.1$ s. For comparison, in the scheme of Ref.~\cite{munro2}, a rate of 2500 pairs per second is achieved for $L=1000$ km and final fidelities higher than $F=0.99$, requiring around 90 qubits per repeater station. Roughly, this corresponds to a rate of 55 pairs per second per memory. However, in that scheme, the fundamental distance has a different value, $L_0=40$ km, and those authors assumed perfect local gates, making the comparison not completely fair. 

The original encoded repeater of Ref.~\cite{jiang} achieves a generation rate of 100 pairs per second for long distances ($L>1000$ km) with final fidelities of $F=0.9984$. However, the system parameters used in that analysis
are quite different from those presented here. In Ref.~\cite{jiang}, the fundamental distance is $L_0=10$ km, the decoherence time is $\tau_c\approx7$ ms, the effective error parameter is $q_{eff}=0.3\%$, and approximately $6n$ qubits at each station are employed, of which $2n$ are memory qubits, and the $4n$ remaining qubits are employed for the local operations on the memory qubits for QEC. This leads, for example, for a three-repetition code, to a rate of approximately 33 pairs per second per memory. 

We showed here that the problem of imperfect memories can be circumvented if we allow for a large number of initial resources and sufficiently good local gates. We further demonstrated that there are trade-offs between the code's efficiency, the decoherence time, and the local gate errors. Depending on these values, we conclude that QEC codes will not always help, and every single code will be efficient in a different regime. Our HQR with encoding using the Golay code $[23,1,7]$ can achieve rates of 1000 bits/s over 1280 km with final fidelities of about $F=0.95$ provided we have 166 memory qubits per half node of the repeater station with decoherence times of 100 ms. This decoherence time has been already exceeded by one order of magnitude in current experiments using nuclear spins systems.

\subsection*{Acknowledgments}
We thank Bill Munro for useful comments. Support from the Emmy Noether Program of the Deutsche Forschungsgemeinschaft is gratefully acknowledged. In addition, we thank the BMBF for support through the QuOReP program.


\appendix

\section*{Appendix A}

Imagine we want to purify an entangled state from two initial states between repeater stations $\mathbf{A}$ and $\mathbf{B}$, $\rho_{\mathbf{A}_1\mathbf{B}_1}\otimes \rho_{\mathbf{A}_2\mathbf{B}_2}$. Let us consider that the initial states, $\rho_{\mathbf{A}_1\mathbf{B}_1}$ and $\rho_{\mathbf{A}_2\mathbf{B}_2}$, are of the form $A\ket{\phi^+}\bra{\phi^+}+B\ket{\phi^-}\bra{\phi^-}+C\ket{\psi^+}\bra{\psi^+}+D\ket{\psi^-}\bra{\psi^-}$, where for the present purpose, $A$, $B$, $C$, and $D$ are simply constants. Following the purification protocol from Refs.~\cite{bennett, deutsch} and considering the error model from Eq.~(\ref{gateerror}), the resulting (unnormalized) state $\rho_c$ is
\begin{align}
\rho_{c}&=(1-q_g(x))^4((A^2+D^2)\ket{\phi^+}\bra{\phi^+}+2AD\ket{\phi^-}\bra{\phi^-}\nonumber\\
&+(B^2+C^2)\ket{\psi^+}\bra{\psi^+}+2BC\ket{\psi^-}\bra{\psi^-})+... .
\label{rhoc}
\end{align}
The terms represented by (...) are those where at least one error occurred in the two-qubit gates. Note that for the case without encoding, these terms can be easily calculated. However, with encoding, especially for large codes, the explicit derivation of these terms is extremely complicated.

The final fidelity and the probability of success of purification are given by
\be
F_{pur}=\frac{\bra{\phi^+}\rho_{c}\ket{\phi^+}}{\rm{Tr} \rho_{c}},\quad\textrm{and}
\ee
\be
P_{pur}=\rm{Tr} \rho_{c}.
\ee
Since we do not know the exact form of $\rho_c$, we will estimate these quantities in a worst-case scenario, thus aiming at lower bounds. For the fidelity, one such bound is obtained when the denominator of the fraction takes on its maximum value and the numerator is just given by the corresponding terms explicitly shown in Eq.~(\ref{rhoc}), resulting in
\be
F_{pur,lower}=\frac{(A^2+D^2)(1-q_g(x))^4}{(A+D)^2+(B+C)^2}=A'_{pur}(1-q_g(x))^4.
\ee
The denominator was calculated, assuming $q_g\ll 1$, such that in first order of $q_g$, the trace is written as
\begin{align}
\rm{Tr} \rho_{c}=&(1-4q_g(x))((A+D)^2+(B+C)^2)+4q_g(x)\rm{Tr}\rho_?\nonumber\\
\leq&(1-((A+D)^2+(B+C)^2))4q_g(x)+\nonumber\\
&((A+D)^2+(B+C)^2),
\label{trace}
\end{align}
where $\rho_?$ is the term we do not know in Eq.~(\ref{rhoc}) up to coefficients with dominating order $4q_g(x)(1-q_g(x))^3\approx 4q_g(x)$. This corresponds to the probability that one error occurred in one of the two qubits at side $\mathbf{A}$ or $\mathbf{B}$. The inequality appears assuming $\rm{Tr}\rho_?\leq 1$. We showed that $(1-((A+D)^2+(B+C)^2))4q_g(x)+((A+D)^2+(B+C)^2)$ is an upper bound for the denominator in first order of $q_g(x)$. We approximate this bound $(1-((A+D)^2+(B+C)^2))4q_g(x)+((A+D)^2+(B+C)^2)\approx (A+D)^2+(B+C)^2$, assuming that $(1-((A+D)^2+(B+C)^2))\sim q_g(x)$ such that the first term of the sum can again be neglected. Notice also that the numerical values in our rate analysis are not noticeably changed by using this approximation whenever $1-T\leq 0.1\%$. Comparing with the exact formula for one round of purification and imperfect quantum gates in Eq.~(\ref{ppur}) in Appendix C, we can see that this is indeed an upper bound for $\rm{Tr} \rho_{c}$ in first order of $q_g(x)$.

Similarly, we obtain as a lower bound for the probability of success for purification
\be
P_{pur,lower}=((A+D)^2+(B+C)^2)(1-q_g(x))^4.
\ee
Provided that the gate errors are sufficiently small, this bound represents a good estimate of the exact value.

The argument used for approximating the fidelity for the swapping is very similar to that given above. However, an important difference is that the swapping operation is a trace-preserving operation, such that, including gate errors, we can guarantee that the probability of success of swapping will always be 1.

For the encoded state, the number of two-qubit gates necessary to realize the swapping is equal to the number of physical qubits per block, $n$. This is the reason why in Eq.~(\ref{Fswap}) the fidelity is multiplied by a factor of $(1-q_g(x))^{2n}$. The same explanation applies to the purification step, but in this case, we obtain a factor of $(1-q_g(x))^{4n}$; here a two-qubit gate has to be applied to each qubit of every entangled pair.

For more rounds of purification, the same pattern is followed. Considering sufficiently many initial spatial resources, for the $k$th-round of purification ($k\geq 1$), the lower bound will be
\begin{align}
P_{pur,lower,k}=&(P_{pur}(\underbrace{A'_{pur}(...A'_{pur}}_{(k-1)-\text{times}}(A,B,C,D))))...(P_{pur}(A,B,C,D))\nonumber\\
&\times(1-q_g(x))^{4n(2^k-1)},
\label{Ppur.rep}
\end{align} 
using Eqs.~(\ref{Fpure}, \ref{Ppure}). For the total fidelity, after the $k$th-round of purification and $N-1$ connections, the lower bound is given by
\begin{widetext}
\begin{equation}\label{Ftotal1}
\underbrace{A'_{swap}(...A'_{swap}(}_{( \log_2{N})-\text{times}}\underbrace{A'_{pur}(...A'_{pur}}_{k-\text{times}}(A_{eff}(F,t_k),B_{eff}(F,t_k),C_{eff}(F,t_k),D_{eff}(F,t_k)))))(1-q_g(x))^{2n((N-1)+2(2^k-1))},
\end{equation}
\end{widetext}
using Eqs.~(\ref{Fpure}-\ref{Fswape}). 

\section*{Appendix B}

Similarly to what was presented for the three-repetition code in Sec.~\ref{hqr.rep}, the state $\frac{\ket{\bar{0}}+\ket{\bar{1}}}{2}$ for the five-repetition code can be deterministically obtained by an interaction of the qubus with the atomic qubits described by $U_{int}^1\left(\theta\right)U_{int}^2\left(2\theta\right)U_{int}^3\left(4\theta\right)U_{int}^4\left(8\theta\right)U_{int}^5\left(-15\theta\right)$ and an $x$ quadrature measurement on the qubus. Note that depending on the measured value of $x$, a phase shift and local bit flip operations may still be applied to change the resulting state to the desired one. In a more systematic way, for an $n$-repetition code, this interaction sequence can be written as $\prod_{j=1}^{n-1}U_{int}^j\left(2^{j-1}\theta\right)U_{int}^n\left(-(2^{n-1}-1)\theta\right)$. Assuming that we want to distinguish all the $\ket{\beta e^{\pm i\theta_j}}$  rotated components for different $j$'s, $(2^{n-1}-1)$ must not be bigger than $\pi$. For $\theta\sim 10^{-2}$, this requirement is not fulfilled for codes with $n\geq 11$, where already for $n=11$, $(2^{10}-1)\theta\sim 3\pi $.

\begin{figure}[t]
\begin{center}
\epsfig{file=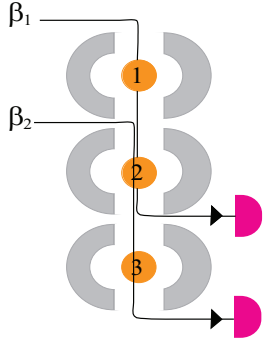,scale=0.45}
\end{center}
\caption{(Color online) Preparing the state $\ket{\bar{0}}+\ket{\bar{1}}$. The normalization factor is omitted. The qubits are initiated in the state $\left(\ket{0}+\ket{1}\right)^{\otimes 3}$. First, a qubus $\ket{\beta_1}$ interacts with atomic qubits 1 and 2. Then a second qubus $\ket{\beta_2}$ interacts with qubits 2 and 3. Both qubuses have their $x$ quadrature measured. All results are valid, since the generation is deterministic. Depending on the measurement results, a phase shift and local bit-flip operations should be applied to the resulting qubit state.}
\label{app} 
\end{figure}

An alternative scheme uses more qubuses for these interactions. Let us start with the three-qubit repetition code again. The encoded state $\frac{\ket{\bar{0}}+\ket{\bar{1}}}{2}$ is generated as illustrated in Fig.~\ref{app}. First, the qubus $\ket{\beta_1}$ interacts with the atoms placed in cavities 1 and 2, with interactions described by
\begin{align}
&U_{int}^1\left(\theta\right)U_{int}^2\left(-\theta\right)\left[\left(\frac{\ket{0}+\ket{1}}{\sqrt{2}}\right)^{\otimes 2}\ket{\beta_1}\right]=\nonumber\\
&\frac{1}{2}\left[\left(\ket{00}+\ket{11}\right)\ket{\beta_1}+\ket{01}\ket{\beta_1 e^{i \theta}}+\ket{10}\ket{\beta_1 e^{-i \theta}}\right].
\end{align}
Then a second qubus $\ket{\beta_2}$ interacts with the atoms placed in cavities 2 and 3 as follows,
\begin{widetext}
\begin{align}
&U_{int}^2\left(-\theta\right)U_{int}^3\left(\theta\right)\left[\frac{\left(\ket{00}+\ket{11}\right)\ket{\beta_1}+\ket{01}\ket{\beta_1 e^{i \theta}}+\ket{10}\ket{\beta_1 e^{-i \theta}}}{2}\left(\frac{\ket{0}+\ket{1}}{\sqrt{2}}\right)\ket{\beta_2}\right]=\nonumber\\
&\frac{1}{2\sqrt{2}}\left[\left(\ket{000}+\ket{111}\right)\ket{\beta_1,\beta_2}+\ket{001}\ket{\beta_1,\beta_2e^{-i\theta}}+\ket{010}\ket{\beta_1e^{i\theta},\beta_2e^{i\theta}}+\ket{100}\ket{\beta_1 e^{-i \theta},\beta_2}\right.\nonumber\\
&\left.+\ket{110}\ket{\beta_1,\beta_2e^{i \theta}}+\ket{101}\ket{\beta_1e^{-i \theta},\beta_2e^{-i \theta}}+\ket{011}\ket{\beta_1e^{i \theta},\beta_2}\right].
\end{align}
\end{widetext}
By measuring the $x$ quadrature separately for each of the two qubuses ($\ket{\beta_1}$ and $\ket{\beta_2}$), the encoded state is deterministically produced. For larger codes, the same procedure can be applied: always alternate $\theta/2$-rotations with $-\theta/2$-rotations and use $n-1$ qubuses interacting only with one pair of atoms of the $n$-qubit chain. Finally, each atom, ignoring the atoms at the ends of the chain, interacts with two different qubus states.

Note that, similarly to Ref.~\cite{louis2}, the scheme proposed here could also be used to generate cluster states via weak nonlinearities. However, in Ref.~\cite{louis2}, the cluster states are obtained through homodyne measurements in the $p$ quadrature of the qubus, which makes that scheme probabilistic.

\section*{Appendix C}

The effective error probability $q_{eff}$ estimates the probability that a physical qubit suffers an odd number of $Z$ errors. In fact, assuming these probabilities sufficiently small \cite{jiang}, it estimates the probability that each physical qubit suffers one $Z$ error. The effective error probability depends on the error parameters ($(1-F)$, $q_g(x)$, and $q_m(t)$) introduced through Eqs.~(\ref{finalstate1}, \ref{gateerror}, \ref{memory.dephasing.2}). We choose the $Z$-error, because in our scheme it occurs more frequently than the $X$-error. 

Before calculating the effective error probability, we should examine the effect of a CNOT gate. An error that initially affects only the target qubit will, after the CNOT operation, also result in an error on the control qubit, in such a way that these errors (and their probabilities) will accumulate.

The first step for the CSS-encoding protocol presented here is the entanglement creation (and eventually purification of the entangled states). For this, the probability that each physical qubit suffers a phase flip is given by $q_1=(1-F_k)+q_m(t/2)$, where $k$ is the number of rounds of purification. For $k=0$, $F_0$ is simply the initial fidelity $F$. After the entangled, possibly purified, pairs have been created, the logical qubits are locally prepared and each physical qubit is subject to a $Z$-error probability of $q_2=q_m(t/2)$. After this, the encoded entangled state is generated by teleportation-based CNOT gates, and the errors accumulate, such that the error probabilities are $q_{3,c}=q_1+q_2+q_g(x)$ and $q_{3,t}=q_2$, for control and target qubits, respectively. After entanglement connections take place, the accumulated probability for obtaining a wrong output is $q_{4,c}=q_{3,c}+q_{3,t}+q_g(x)$ and $q_{4,t}=q_2$. For simplicity, we may just use the largest from these two values to estimate the effective error probability per physical qubit, such that
\be
q_{eff}=3q_m(t/2)+(1-F_k)+2q_g(x).
\label{qef}
\ee
The decaying time $t$ is considered to be the time it takes for classical communication to announce that entanglement distribution succeeded ($T_0/2$) and the time it takes to announce that purification succeeded (again $T_0/2$). Hence $t=t'_k=(k+1)T_0/2$. Note that $t'_k$ is different from the decaying time $t_k$ for the repetition code by $T_0/2$. The reason for this is that for the repetition code protocol, the logical qubits are already decaying from the very beginning. We should be careful in defining $F_k$, because gate errors must be included here. If we start with two copies of the entangled state $A\ket{\phi^+}\bra{\phi^+}+B\ket{\phi^-}\bra{\phi^-}+C\ket{\psi^+}\bra{\psi^+}+D\ket{\psi^-}\bra{\psi^-}$, following the purification protocol from Ref.~\cite{deutsch} and considering the gate error model from Eq.~(\ref{gateerror}),the resulting state after one round of purification is given by $A'\ket{\phi^+}\bra{\phi^+}+B'\ket{\phi^-}\bra{\phi^-}+C'\ket{\psi^+}\bra{\psi^+}+D'\ket{\psi^-}\bra{\psi^-}$, where
\begin{widetext}
\begin{align}
A'=\frac{1}{P_{pur}^{imp}}&\left(D^2 + A^2 (1 + 2 (-1 + q_g) q_g)^2 - 
 2 A (-1 + q_g) q_g (C + 2 D + 2 (B - C - 2 D) q_g + 2 (-B + C + 2 D) q_g^2)\right.\nonumber\\
     &\left.-2 D (-1 + q_g) q_g (-2 D - 2 (C + D) (-1 + q_g) q_g + B (1 + 2 (-1 + q_g) q_g))\right)\nonumber\\
B'=\frac{1}{P_{pur}^{imp}}&\left(-2 D (-1 + q_g) q_g (C + D - 2 (-B + C + D) q_g + 2 (-B + C + D) q_g^2) + 
 2 A^2 q_g (1 + q_g (-3 - 2 (-2 + q_g) q_g)) +\right.\nonumber\\
     &\left. 2 A (D (1 + 2 (-1 + q_g) q_g)^2 - (-1 + q_g) q_g (-2 C (-1 + q_g) q_g + 
      B (1 + 2 (-1 + q_g) q_g)))\right)\nonumber\\
C'=\frac{1}{P_{pur}^{imp}}&\left(C^2 + B^2 (1 + 2 (-1 + q_g) q_g)^2 - 
 2 C (-1 + q_g) q_g (-2 C - 2 (C + D) (-1 + q_g) q_g + A (1 + 2 (-1 + q_g) q_g))\right.\nonumber\\
 &\left. -2 B (-1 + q_g) q_g (-2 A (-1 + q_g) q_g + D (1 + 2 (-1 + q_g) q_g) + 
   C (2 + 4 (-1 + q_g) q_g))\right)\nonumber\\
D'=\frac{1}{P_{pur}^{imp}}&\left(-2 C (-1 + q_g) q_g (C + D - 2 (-A + C + D) q_g + 2 (-A + C + D) q_g^2) + 
 2 B^2 q_g (1 + q_g (-3 - 2 (-2 + q_g) q_g)) +\right.\nonumber\\
     &\left.  2 B (C (1 + 2 (-1 + q_g) q_g)^2 - (-1 + q_g) q_g (-2 D (-1 + q_g) q_g + 
      A (1 + 2 (-1 + q_g) q_g)))\right),
\label{coeffpur}
\end{align}
and $P_{pur}^{imp}$ is the purification probability of success given by
\be
P_{pur}^{imp}=(B + C)^2 + (A + D)^2 - 2 (A - B - C + D)^2 q_g + 
 2 (A - B - C + D)^2 q_g^2.
\label{ppur}
\ee
For the case of $q_g=0$, Eqs.~(\ref{coeffpur}, \ref{ppur}) are in accordance with Ref.~\cite{deutsch}.
\end{widetext}
The fidelity after the first round of purification, $F_1$, is given by $A'$ when $A=F$, $B=1-F$, and $C=D=0$. For small $q_g$ and high initial fidelity, which is the regime under consideration here, the dominant coefficient (after $A'$) is $B'$. Note that $B'\approx(1-F_1)$ for $q_g\ll 1$, and thus the probability that one physical qubit suffers an error becomes $(1-F_1)$. For more rounds of purification, a similar procedure can be performed.

We considered the probability of no error per physical qubit immediately after one round of purification as $(1-q_m(t/2))F_1$, such that the probability of one error is approximated by $(1-F_1)+q_m(t/2)$. The purification protocol, however, can improve the fidelity against memory dephasing that happened during the entanglement distribution. This can be computed by calculating $F_1$, substituting $A=F(1-q_m(t/2))+(1-F)q_m(t/2)$, $B=(1-F)(1-q_m(t/2))+F q_m(t/2)$, and $C=D=0$ in $A'$, with $t=T_0/2$. Although this strategy can improve the final fidelity, the qubits decay further after the purification step, and so, for simplicity, we shall ignore this fact. Indeed, in this case, the probability of success for the purification should be smaller, however, for small probabilities of errors, this difference is so small and we may neglect it. 

We should notice here that Eq.~(\ref{qef}) is not identical, though it is similar, to the one presented in App. A of Ref.~\cite{jiang}. This lies in the fact that, although our protocol was inspired by the paper of Jiang \textit{et al.}, there are some crucial differences. To cite one, in our analysis, we do not assume that our purified entangled pairs are perfect, and the imperfect generation of an entangled pair is also included as an error. In addition, the qubits suffer memory dephasing errors already during the purification step. Finally, our error model is different from that used in Ref.~\cite{jiang}.

\end{document}